\documentstyle[11pt,newpasp,twoside,epsf]{article}
\def\edcomment#1{\iffalse\marginpar{\raggedright\sl#1\/}\else\relax\fi}
\reversemarginpar

\newcommand{\3}{\mbox{$^{-3}$}}

\newcommand{\MPC}{\mbox{${\rm \,Mpc}$}}

\newcommand{\zo}{\mbox{$z_\circ$}}
\newcommand{\tzo}{\mbox{$t_\circ$}}
\newcommand{\NBH}{\mbox{$N(M,t)$}}
\newcommand{\NAGN}{\mbox{$\phi(L,t)$}}

\newcommand{\Mdot}{\mbox{$\dot{M}$}}
\newcommand{\MdotBH}{\mbox{$\Mdot(M,t)$}}
\newcommand{\rhobh}{\mbox{ $\rho_{\rm BH}$ }}
\newcommand{\RAGN}{\mbox{${\rm R}_{2-1}$}}
\newcommand{\RAGNB}{\mbox{${\rm R}_{2-1,B}$}}
\newcommand{\RAGNX}{\mbox{${\rm R}_{2-1,X}$}}
\newcommand{\etal}{et al.\ }
\newcommand{\Msun}{\mbox{${\rm M}_\odot$}}

\newcommand{\Mrat}{\mbox{$x$}}
\newcommand{\MBH}{\mbox{$M_{\rm BH}$}}
\newcommand{\dlnL}{\mbox{${\rm d\ln}L$}}
\newcommand{\dM}{\mbox{${\rm d}M$}}

\newcommand{\MBulge}{\mbox{$M_{\rm Bulge}$}}

\newcommand{\nuB}{\mbox{$\nu_{\rm B}$}}
\newcommand{\xten}[1]{\mbox{$\times 10^{#1}$}}
\newcommand{\ten}[1]{\mbox{$10^{#1}$}}

\begin{document}
\title{Relic Black Holes and Type 2 Quasars}
 \author{Alessandro Marconi and Marco Salvati}
\affil{Osservatorio Astrofisico di Arcetri, Largo Fermi 5, 50125 Firenze, Italy}

\begin{abstract}
We follow the evolution of the Black Hole mass function in the assumption that
mass accretion onto a massive BH is the powering mechanism of AGN activity.
With reasonable assumptions suggested by current knowledge, it is possible to
reproduce the Black Hole mass function derived from local bulges. In particular
it is found that the majority of the mass in relic Black Holes was produced
during AGN activity and that the ratio between obscured and unobscured quasars
is between 1 and 2, much lower than required by synthesis models of the X-ray
background.
\end{abstract}

\section{Introduction}

It has long been suspected that the most luminous Active Galactic Nuclei (AGN)
are powered by accretion of matter onto massive black holes (BH).
This belief, combined with the observed evolution of the
space-density of AGN suggests that massive BHs should be present in local
galaxies as relics of past activity.  Indeed, the current observational
evidence indicates that most, possibly all, luminous galaxies host a massive BH
at their centres and its mass correlates with the luminosity and stellar
velocity dispersion of the bulge in which it resides (e.g. Richstone \etal
1998, Ferrarese \& Merritt 2000).
Soltan (1982) and
Chokshi \& Turner (1992) derived the local mass density in relic BHs, \rhobh,
expected from past activity using optical number counts of AGNs.  More
recently, Fabian \& Iwasawa (1999) estimated \rhobh\ from the energy density of
the hard X-Ray Background (XRB) and this value is in agreement with the {\it
observed} local BH density derived by Merritt \& Ferrarese (2001) using the
BH-mass vs Bulge-mass relation.  However it seems that the AGNs found in
optical and soft X-ray surveys are not enough to account for the local \rhobh\
and this fact prompted Fabian \& Iwasawa to talk about "obscured accretion",
i.e. obscured AGN activity.
The existence of obscured AGNs (the so-called "type 2" objects) is required to
account for the XRB spectrum with integrated emission from AGNs (e.g. Gilli,
Salvati \& Hasinger 2001 and references therein): the ratio between type 2 and
unobscured type 1 objects (hereafter \RAGN) is usually in the range 4-10,
implying that both optical and Xray surveys are missing a large fraction of the
AGN population.  With a completely independent argument, the missing Lyman edge
in blue quasars implies that \RAGN\ is at least $\ge 0.5$ (Maiolino \etal
2001).  Locally, low luminosity obscured AGNs are well known (Seyfert 2
galaxies) and in the ratio of about 4:1 with respect to type 1 objects
(Maiolino \& Rieke 1995).  However, at higher redshifts few type 2 quasars
(high luminosity objects) are known, a number which is largely insufficient to
account for the spectral shape of the XRB.  What is the relationship between
type 2 AGNs and massive black holes?  The arguments based on \rhobh\ indicates
that some obscured AGNs must exist in order to account for the fossil mass but
nothing can be told about the luminosity of these obscured objects: are many
low luminosity AGNs needed or could fewer high luminosity quasars suffice?  Is
it really true that AGN activity can explain the local BHs, not only globally,
but also in terms of their mass distribution?  This paper will try to answer
these questions by comparing the local BH mass function (BHMF) expected from
AGN activity with the local BHMF inferred from galaxy bulges. Throughout this
paper we assume $H_\circ=70$km/s/Mpc$^{-1}$, $\Omega_{\rm M}=1$ and
$\Omega_\Lambda=0$.

\begin{figure}
   \plottwo{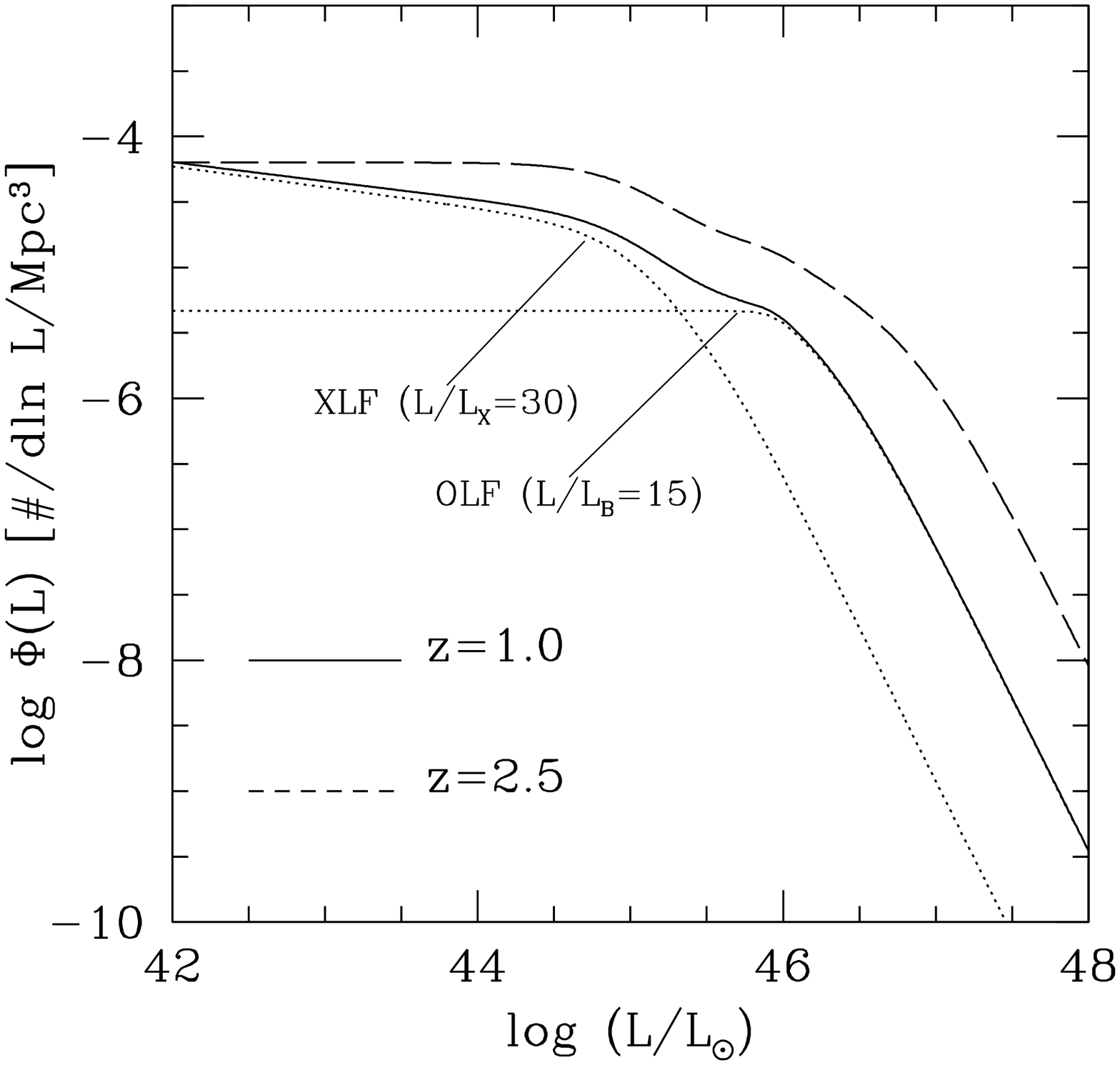}{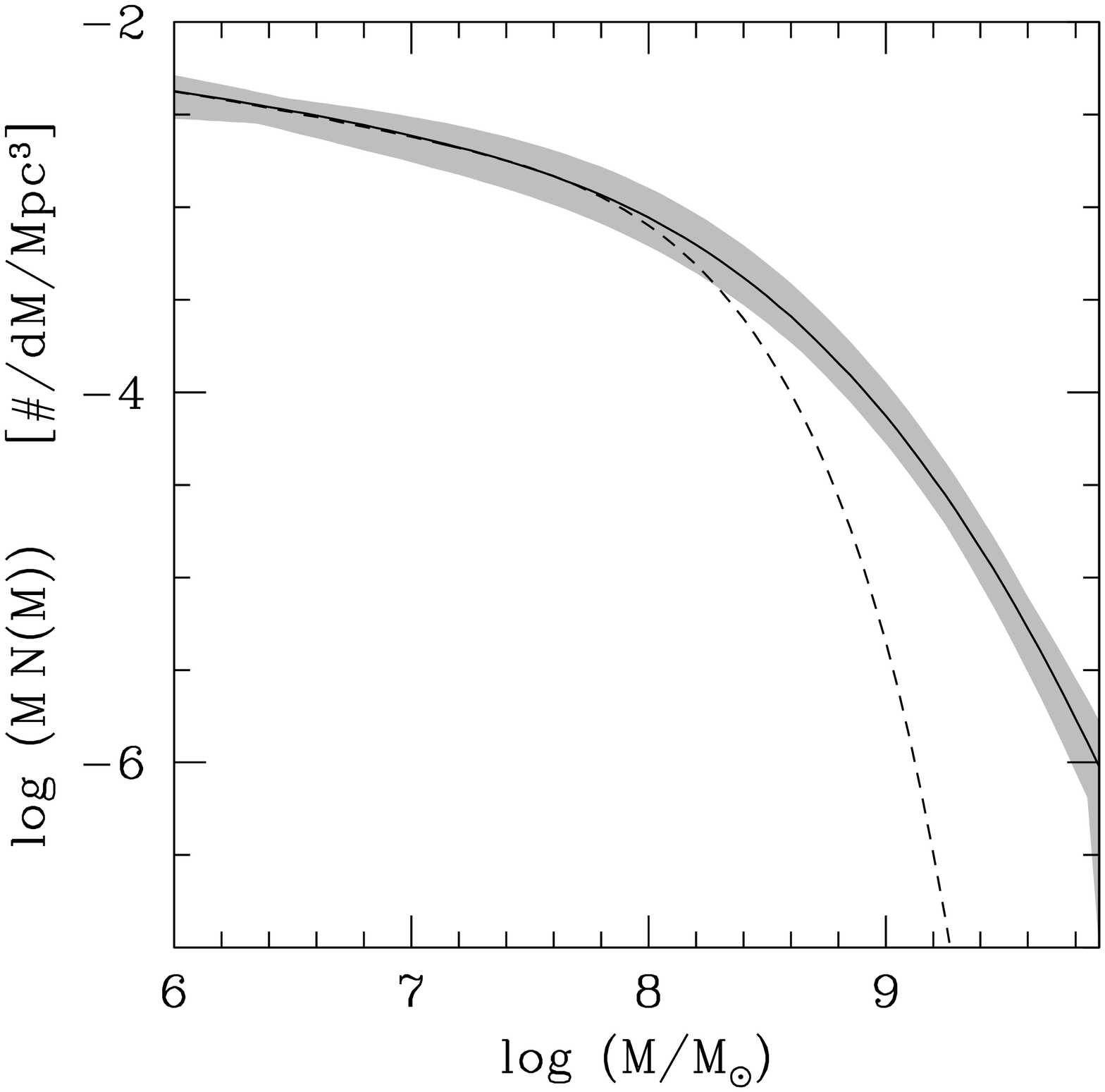}
   \caption{\label{fig:AGNLF}(a) AGN luminosity function at $z=1.0$ (solid line)
   and $z=2.5$ (dashed line).
   This is obtained by combining the XLF by Gilli, Salvati \& Hasinger 2001
   (low $L$) and the OLF by Boyle \etal 2000 (high $L$).
   \label{fig:BHMF}(b) Local BH Mass Function (solid line).  The
   uncertainties are represented by the gray area.  The dashed line is the BHMF
   obtained assuming that $\sigma_{\Mrat}=0$.  }
\end{figure}

\section{The continuity equation}

We consider the continuity equation which regulates the time evolution of the
BH mass function with the Small \& Blandford (1992) formalism. If \NBH\ is the
BH mass function ($\NBH {\rm d}M$ is the number of BHs per unit comoving volume
(Mpc$^{-3}$) at cosmic time $t$), the continuity equation can be written as:
\begin{equation}\label{eq:cont} \frac{\partial\NBH}{\partial t} +
\frac{\partial}{\partial M} \left(\NBH <\MdotBH> \right) = S(M,t)
\end{equation} where $<\MdotBH>$ is the "average" accretion rate and $S(M,t)$
is the source function.  As in Small \& Blandford (1992) we assume that a
massive BH is "born" when its mass is $M\ge M_{\rm min}$ and that no BH is
created or destroyed (i.e. we neglect any merging process). With these
assumptions the source function is simply $S(M,t) = S_\circ \delta(M-M_{\rm
min})$ where $\delta$ is the Dirac delta function.
If AGNs are powered by accretion onto BHs, we can relate the AGN Luminosity
Function $\phi(L,t)$ ($\NAGN {\rm dln} L$ is the number of AGNs 
per unit comoving volume (Mpc$^{-3}$) at
cosmic time $t$) to the BH mass function 
\begin{equation}\label{eq:AGNBH}
\phi(L,t) \dlnL = \delta(M,t) N(M,t) \dM
\end{equation}
where $\delta(M,t)$ is the fraction of BHs with mass M active at time $t$.
If a BH is accreting at a fraction
$\lambda$ of the Eddington rate its emitted luminosity is then
$L=\lambda Mc^2/t_E = \varepsilon \Mdot c^2$
where $t_E$ is the Eddington time, 
$\varepsilon$ is the accretion efficiency and \Mdot\ is the accretion rate.
Since $<\MdotBH> = \delta(M,t) \MdotBH$, we can write:
\begin{equation}
\NBH <\MdotBH> = \frac{\lambda}{\varepsilon t_E}
	       		\left[ \phi(L,t) \right]_{L=\lambda\frac{Mc^2}{t_E}} 
\end{equation}
\begin{equation}
\frac{\partial\NBH}{\partial t} = -\frac{c^2\lambda^2}{\varepsilon t_E^2}
\left[ \frac{\partial\phi(L,t)}{\partial L} \right]_{L=\lambda\frac{Mc^2}{t_E}}
\end{equation}
which can be easily integrated given the AGN luminosity function AND the initial
conditions.
When $\lambda$ is dependent on $M$ the equation is more complex but
still easily solvable.
As initial conditions, we assume that at a given redshift \zo\ ($\tzo=t(\zo)$)
ALL Black Holes are active, i.e. that $\delta(M,t(\zo))=1$.
The only constraint is that $\delta(M,t) = \phi(L,t)/MN(M,t) \le 1$.

A fundamental ingredient is the AGN luminosity function (LF) expressed in terms
of the total AGN luminosity, $L$.  We first consider the LF of optically
selected quasars (OLF in the B band; Boyle \etal 2000) and the LF of soft X-ray
selected AGNs (XLF in 0.5-2 keV band; Gilli \etal 2001, it represents only
unabsorbed AGNs and is derived using the XLF by Miyaji, Hasinger \& Schmidt
2000). Both LFs consider only type 1 (unobscured) AGNs.  In principle the LF in
one spectral band should be enough to obtain $\phi(L,t)$ but in this paper we
consider more instructive to combine the above LFs using Boyle's LF to describe
high $L$ AGNs and Gilli's one to describe low $L$ AGNs.  A more refined
treatment will be presented elsewhere.
The AGN luminosity function for type 1 objects is plotted in Fig.
\ref{fig:AGNLF}a at two redshift values. We have assumed $L/L_{\rm X}=30$,
for low L objects and $L/\nuB L_{\nuB}=15$
as typical of quasars (e.g. Elvis \etal 1994).

\begin{figure}
   \plottwo{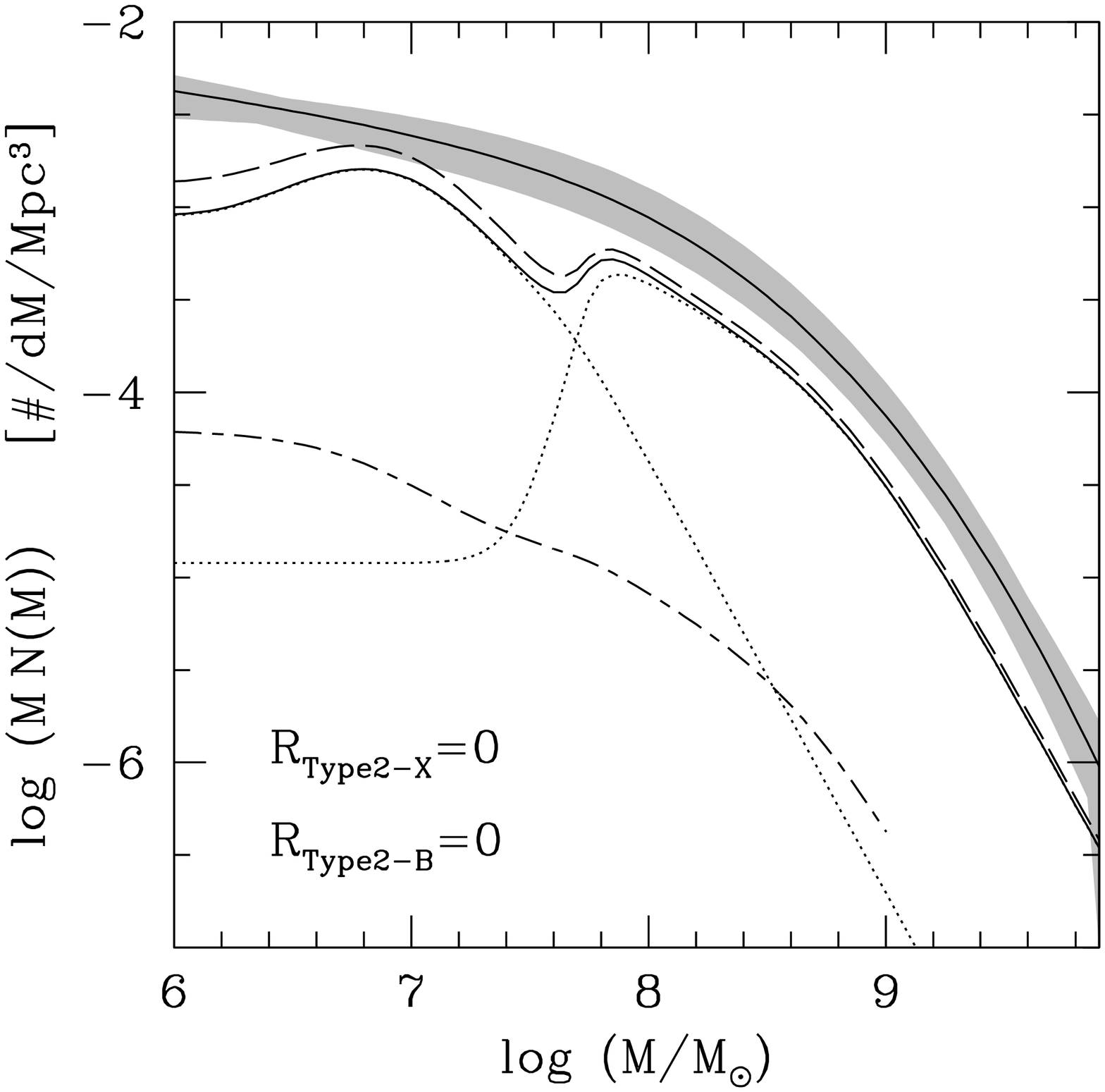}{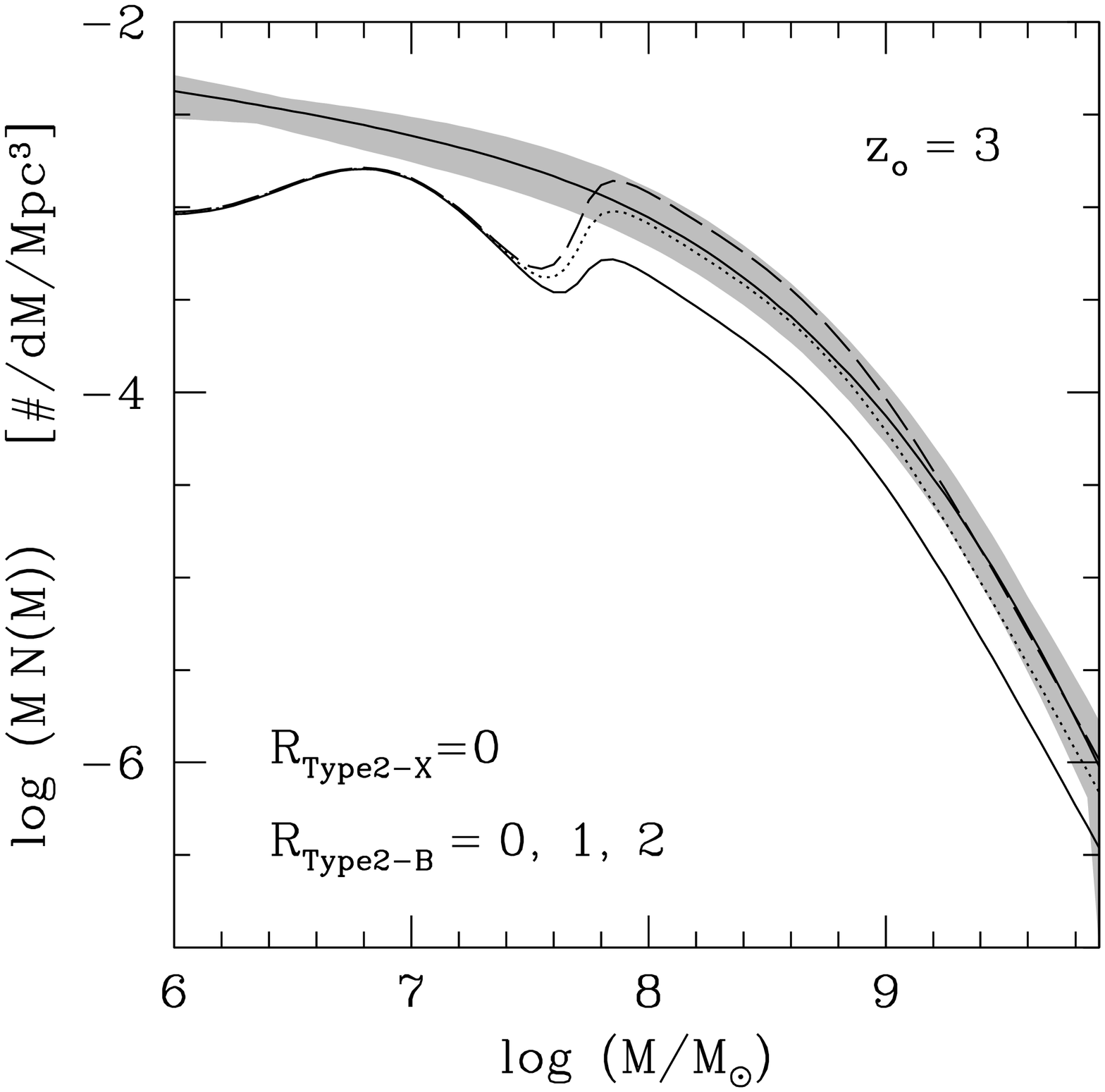}
   \caption{\label{fig:m01}(a) Local BH mass function from galaxy bulges (solid
   line in grey area) compared with BH mass function obtained considering only
   type 1 AGNs (solid line). The dotted lines are the 	contributions from low
   and high $L$ AGNs. The dot-dashed line is the BH mass function at the
   redshift \zo=3 when all BHs were active. The dashed line represents the BHMF
   from AGN activity assuming that only 10\% of the BHs were
   active at \zo=5.
   \label{fig:m02}(b)The contribution
   from high $L$ AGNs is computed for \RAGNB=0 (solid line), 1 (dotted line), 2
   (dashed line).}
\end{figure}
\begin{figure}
   \plottwo{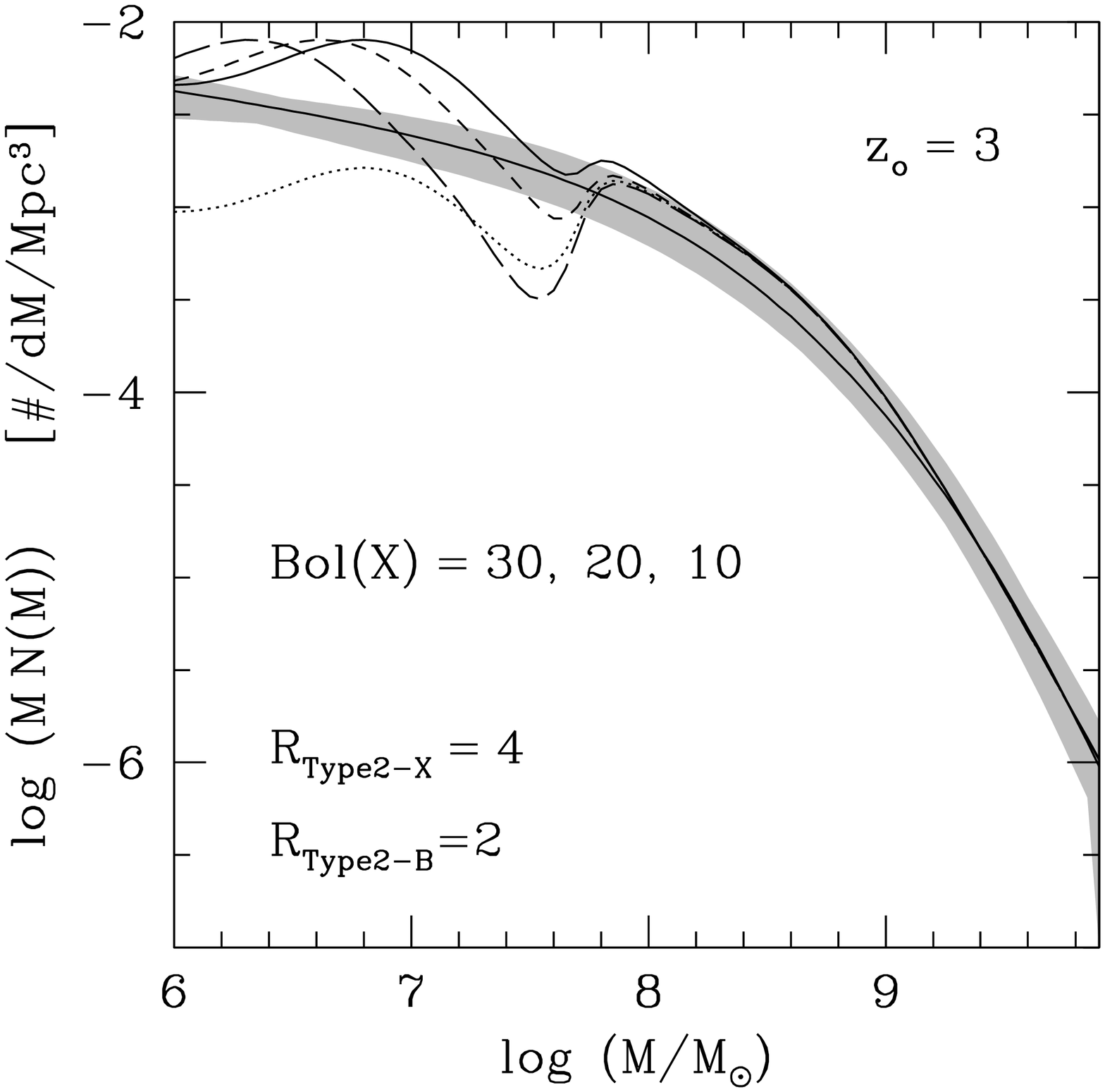}{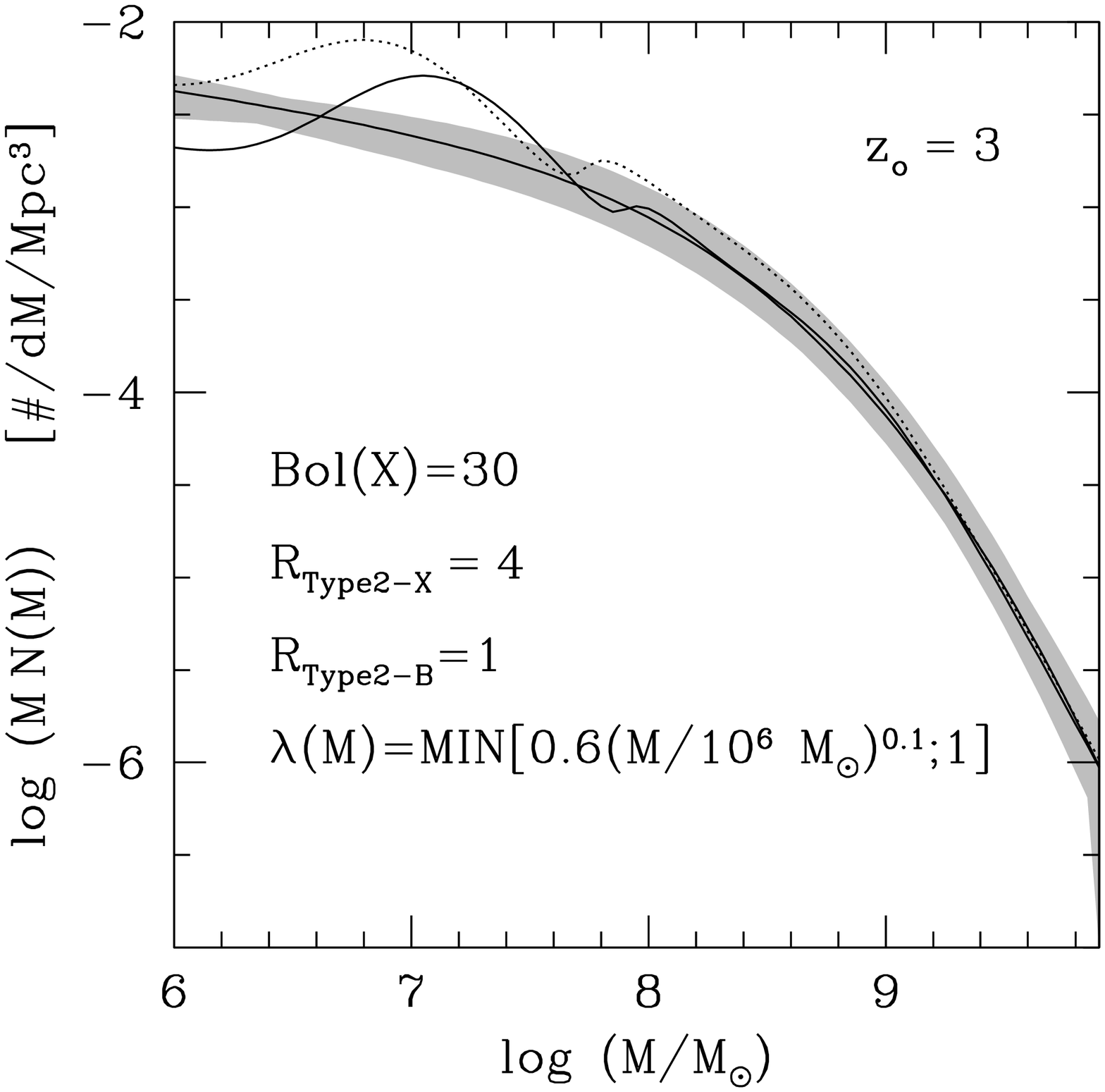}
   \caption{\label{fig:m03}(a) BHMF from AGN activity computed with
   \RAGNX=4 for $L/L_{\rm X}=$ 30 (solid
   line), 20 (short-dashed line), 10 (long-dashed line).  The dotted line
   represent the model with \RAGNX=0.  \label{fig:m04}(b) The dotted and solid
   lines represent models with $L/L_{\rm X}=30$, $L/\nuB L_{\nuB}=15$
   and \RAGNX=4. In the model represented by the dotted line all AGNs
   are emitting at their Eddington luminosity, while with the solid line AGNs
   are emitting at a fraction of their Eddington luminosity which depends on
   $M$ as specified in the figure.  For this model \RAGNB=1.}
\end{figure}

\section{The local BH mass function}

To estimate the local BH mass function we follow the method by Salucci \etal
(1999).  The BHMF is derived from the Galaxy LF through the following steps:
(1) we consider the LF of galaxies in the B band divided by morphological type
(Folkes \etal 1999); (2) for each morphological type we
derive the LF of bulges applying the bulge/disk ratio by Simien \& de
Vaucouleurs (1985); (3) we obtain the mass function of bulges using the
mass-to-light ratio by Fukugita, Hogan, \& Peebles (1998); (4) we convolve the
bulge mass function with the probability distribution of the ratio
$\MBH/\MBulge$ to obtain the BH mass function. An important assumption in the
last point is that the observed scatter in the \MBH - \MBulge\ relation is real
and not due to observational uncertainties.  We then assume a gaussian
probability distribution $P(\log {\Mrat})$ ($\Mrat = \MBH/\MBulge$) with
$<\log\Mrat>=-2.93$ and $\sigma_{\Mrat}=0.45$ as derived by Merritt \&
Ferrarese (2001).  The BHMF is presented in Fig. \ref{fig:BHMF}b and the "grey"
region represent an error estimate on the BH mass function obtained by varying
the parameters used in each of the above steps.  The BHMF has $\rhobh=
(5\pm2)\xten{5} \Msun\MPC\3$ and this value is in agreement with $\rhobh=
(3\div 6)\xten{5} \Msun\MPC\3$ obtained by Fabian \& Iwasawa (1999) and Salucci
\etal (1999) from the hard XRB.

\section{Model Results}
Model results are shown in Fig. \ref{fig:m01}a.  The solid line in the grey area
is the BH mass function derived from the galaxy mass function as indicated
in the previous section.
The solid line is the BH mass function expected from type 1 AGNs only with
a starting redshift of $\zo=3$. The starting redshift at which we make the
assumption of $\delta(M,\tzo)=1$ (or alternatively the redshift at which we
select the active BHs whose evolution we will follow) is not important for the
final results.  Indeed the dotted line representing the BHMF obtained assuming
\zo=5 and $\delta=0.1$, 
is not significantly different from the one with \zo=3 and $\delta=1$.
The only
constraint is that $\zo>z_{\rm max}$, the redshift of maximum AGN
activity.  From Fig. \ref{fig:m01}a it is clear that Optical QSOs (high $L$
AGNs) produce black holes with masses $\MBH>\ten{8}\Msun$ while X-ray AGNs (low
$L$ AGNs) produce the lower mass ones.
Moreover, comparing the BHMF obtained from AGN activity at $z=0$ with the one
at \zo\ it can be clearly seen that BH masses are
produced during the AGN activity: this is the main
reason why the results are little sensitive to the initial conditions.
The model presented in the figure has
\RAGN=0, i.e. it considers only unobscured AGNs. Thus it is not unexpected to
derive a low \rhobh=1.8\xten{5}\Msun\MPC\3.
Let's first consider the high $L$ AGNs. It is not yet clear if type 2 quasars
exist or not or what is their ratio \RAGN\ relative to type 1 objects.
The comparison between the "observed" local BHMF with the one deduced 
from AGN activity can yield a constraint on \RAGN.
From Fig. \ref{fig:m02}b,  
$1<\RAGN<2$ is required to match the local BHMF in absolute level (the slope
is nicely reproduced). Correspondingly $\rhobh=3.2\div 4.6 \xten{5}\MBH\MPC\3$.
Let's now focus on the low $L$ AGNs. The existence of low $L$ type 2 AGNs
is well know and the local ratio is \RAGN=4 (e.g. Maiolino \& Rieke 1995).
As shown in Fig. \ref{fig:m03}, with that ratio the local BHMF at low masses
($\MBH<\ten{8}\Msun$) is overproduced.  A change in the bolometric correction
does not improve the situation because the BHMF is just shifted along the M
axis.  A possible solution to this problem is that low luminosity AGNs are not
emitting at the Eddington Luminosity (Fig. \ref{fig:m04}a), a possibility also
considered in other works (e.g. Salucci \etal 1999 and references therein), and
indeed, apart for some "wiggles", the BHMF is now well reproduced.  One must
keep in mind that this might also mean that there are some problems with the
AGN luminosity functions at low luminosities or that the \MBH-\MBulge\ relation
is different at low masses. 
\section{Summary and Conclusions}

We have used the continuity equation to follow the evolution of the BH mass
function in the assumption that mass accretion onto a BH is the powering
mechanism of AGN activity.  AGNs accrete at a fraction of their Eddington
luminosity and at a starting redshift \zo\ all BHs were active.  By comparing
the expected BHMF with the one derived from local galaxy luminosity functions
we can make the following conclusions.  (1) The majority of the mass of a local
BH was produced during AGN activity.  (2) Quasar activity was responsible for
the growth of BHs with $M>\ten{8}\Msun$.  (3) The final results are not
sensitive to the value of the redshift \zo\ at which all BH were active.  (4)
The ratio between type 2 and type 1 quasars is $\RAGN\simeq1-2$, i.e. $\simeq
30\% - 50\%$ of \rhobh\ is made by type 2 QSOs.  (5) The required \RAGNB\ can
be reduced if we assume that $L<L_{Eddington}$ or $\varepsilon<0.1$. In any
case it is clear that (a) \RAGNB\ is much lower than the $4-10$ value required
to fit the XRB and (b) no small efficiencies are required to explain the local
BHMF.  (6) $\RAGNX>0$ gives a problem at low masses overpredicting the number
of relic BHs.  This could indicate that either $L<L_{Eddington}$ at low
luminosities or that there might be uncertainties on the BHMF and/or the AGN LF
at low M and L.  (7) No merging of BHs is required by our model.

The main conclusion of this work is that with "reasonable" assumptions,
compatible with current knowledge, mass accretion onto massive BH during AGN
activity can explain the local BH mass function both in shape and
normalization.  There is not much room for a large ($R>>1$) population of high
luminosity obscured AGNs (i.e. type 2 quasars).

\end{document}